\documentclass[preprints,article,accept,moreauthors,pdftex,10pt,a4paper]{Definitions/mdpi}

\firstpage{1} 
\pubvolume{xx}
\issuenum{1}
\articlenumber{5}
\pubyear{2018}
\copyrightyear{2018}
\history{Received: date; Accepted: date; Published: date}

\Title{Engineering planar transverse domain walls in biaxial magnetic nanostrips by tailoring transverse magnetic fields with uniform orientation}

\Author{Mingna Yu $^{1}$, Mei Li $^{2}$* and Jie Lu $^{1,}$*\orcidA{}}

\address{%
$^{1}$ \quad College of Physics, Hebei Advanced Thin Films Laboratory, Hebei Normal University, Shijiazhuang 050024, Hebei, China; 1543916410@qq.com (M.Y.)\\
$^{2}$ \quad Physics Department, Shijiazhuang University, Shijiazhuang 050035, Hebei, China}

\corres{Correspondence: limeijim@163.com; jlu@hebtu.edu.cn}

\abstract{Designing and realizing various magnetization textures in magnetic nanostructures are essential for developing novel magnetic nanodevices in modern information industry. Among all these textures, planar transverse domain walls (pTDWs) are the simplest and the most basic, which make them popular in device physics. In this work, we report the engineering of pTDWs with arbitrary tilting attitude in biaxial magnetic nanostrips by transverse magnetic field profiles with uniform orientation but tunable strength distribution. Both statics and axial-field-driven dynamics of these pTDWs are analytically investigated. It turns out that for statics these pTDWs are robust again disturbances which are not too abrupt, while for dynamics it can be tailored to acquire higher velocity than Walker's ansatz predicts. These results should provide inspirations for designing magnetic nanodevices with novel one-dimensional magnetization textures, such as 360$^\circ$ walls, or even two-dimensional ones, for example vortices, skyrmions, etc.}


\begin{document}



\section{Introduction}\label{introduction}
Artificially prepared magnetic nanostructures have been forming the basic components of nanodevices in modern information industry for decades\cite{Leeuw_RPP_1980,Bauer_RMP_2005}. Various magnetization textures therein provide the abundant choices of defining zeros and ones in binary world. Among them, domain walls (DWs) are the most common ones which separate magnetic domains with interior magnetization pointing to different directions\cite{XiongG_2005_Science,Parkin_2008_Science_a,Parkin_2008_Science_b,Koopmans_2012_nanotech,Thomas_JAP_2012,Parkin_2015_nonotech}. In magnetic nanostrips with rectangular cross sections, numerical calculations confirm that there exists a critical cross-section area\cite{McMichael_IEEE_1997,Thiaville_JMMM_2005}. Below (above) it, transverse (vortex) walls dominate. For nanodevices based on DW propagation along strip axis with high integral level, strips are thin enough so that only transverse DWs (TDWs) appear. Their velocity under external driving factors (magnetic fields, polarized electronic currents, etc.) determines the response time of nanodevices based on DW propagation. In the past decades, analytical, numerical and experimental investigations on TDW dynamics have been widely performed and commercialized to a great extent\cite{Walker_JAP_1974,Ono_Science_1999,XiongG_nmat_2003,Erskine_nmat_2005,Tretiakov_PRL_2008,Erskine_PRB_2008,jlu_EPL_2009,YanP_AOP_2009,SZZ_PRL_2010,Tatara_JPDAP_2011,Berger_PRB_1996,Slonczewski_JMMM_1996,ZhangSF_PRL_2004,Ono_PRL_2004,Erskine_PRL_2006,Hayashi_PRL_2006,YanP_APL_2010}. However, seeking ways to further increase TDW velocity, thus improve the devices' response performance, is always the pursuit of both physicists and engineers.

Besides velocity, fine manipulations of DW structure are also essential for improving the device performance. In the simplest case, a TDW with uniform azimuthal distribution, which is generally called a planar TDW (pTDW), is of the most importance. Historically the Walker ansatz\cite{Walker_JAP_1974} provides the first example of pTDW, however its tilting attitude is fully controlled by the driving field or current density (in particular, lying within easy plane in the absence of external driving factors) thus can not be freely adjusted. In the past decades, several strategies\cite{Thiaville_nmat_2003,Kim_APL_2007,Bryan_JAP_2008,jlu_JAP_2010} have been proposed to suppress or at least postpone the Walker breakdown thus makes TDWs preserve traveling-wave mode which has a high mobility (velocity versus driving field or current density). The nature of all these proposals is to destroy the two-fold symmetry in the strip cross section, thus is equivalent to a transverse magnetic field (TMF), no matter it's built in or extra. In 2016, the ``velocity-enhancement" effect of uniform TMFs (UTMFs) on TDWs in biaxial nanostrips has been thoroughly investigated\cite{jlu_PRB_2016}. It turns out that UTMFs can considerably boost TDWs' propagation meanwhile inevitably leaving a twisting in their azimuthal planes. However for applications in nanodevices with high density, the twisting is preferred to be erased to minimize magnetization frustrations and other stochastic fields. In 2017, optimized TMF profiles with fixed strength and tunable orientation are proposed to realize pTDWs with arbitrary tilting attitude\cite{limei_srep_2017}. Dynamical analysis on these pTDWs reveals that they can propagate along strip axis with higher velocities than those without TMFs. However, there are several remaining problems: the rigorous analytical pTDW profile (thus TMF distribution) is still lacking, the pTDW width can not be fully controlled and the real experimental setup is challenging. 

In this work, we engineer pTDWs with arbitrary tilting attitude in biaxial magnetic nanostrips by tailoring TMF profiles with uniform orientation but tunable strength distribution. For statics, the well-tailored TMF profile manipulates pTDW with arbitrary tilting attitude, clear boundaries and controllable width. In particular, these pTDWs are robust again disturbances which are not too abrupt. For axial-field-driven dynamics with TMFs comoving, pTDWs will acquire higher velocity than Walker's ansatz predicts.

\section{Model and Preparations}

\begin{figure}[H]
	\centering
	\includegraphics[width=10 cm]{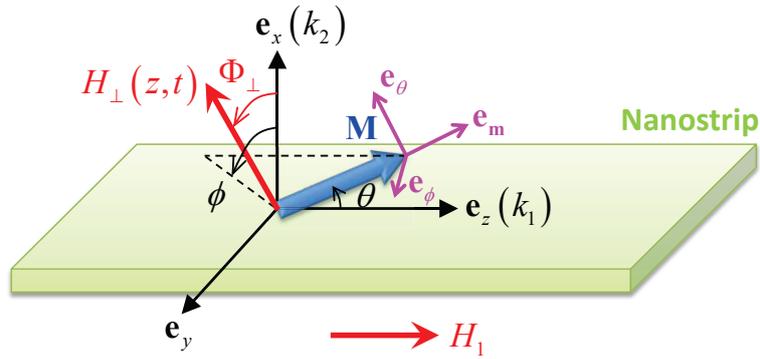}
	\caption{Sketch of biaxial magnetic nanostrip under consideration. ($\mathbf{e}_x,\mathbf{e}_y,\mathbf{e}_z$) is the global Cartesian coordinate system in real space: $\mathbf{e}_z$ is along strip axis, $\mathbf{e}_x$ is in the thickness direction and $\mathbf{e}_y=\mathbf{e}_z\times\mathbf{e}_x$. $k_1(k_2)$ is the total magnetic anisotropy coefficient in easy (hard) axis. ($\mathbf{e}_{\mathbf{m}},\mathbf{e}_{\theta},\mathbf{e}_{\phi}$) forms the local spherical coordinate system associated with the magnetization vector $\mathbf{M}$ (blue arrow with magnitude $M_s$, polar angle $\theta$ and azimuthal angle $\phi$). The total external field has two components: axial driving field with magnitude $H_1$ and TMF with constant tilting attitude $\Phi_{\perp}$ and tunable magnitude $H_{\perp}(z,t)$.}
\end{figure}   

We consider a biaxial magnetic nanostrip with rectangular cross section, as depicted in Figure 1. The $z$ axis is along strip axis, the $x$ axis is in the thickness direction and $\mathbf{e}_y=\mathbf{e}_z\times\mathbf{e}_x$. The magnetic energy density functional of this strip can be written as,

\begin{equation}\label{Energy_Density}
\mathcal{E}_{\mathrm{tot}}[\mathbf{M},\mathbf{H}_{\mathrm{ext}}]=-\mu_0 \mathbf{M}\cdot\mathbf{H}_{\mathrm{ext}}-\frac{k_1}{2}\mu_0 M_z^2+\frac{k_2}{2}\mu_0 M_x^2+J\left(\nabla\mathbf{m}\right)^2,
\end{equation}
in which $\mathbf{m}\equiv \mathbf{M}/M_s$ with $M_s$ being the saturation magnetization. The magnetostatic energy density has been described by quadratic terms of $M_{x,y,z}$ via three average demagnetization factors $D_{x,y,z}$\cite{Aharoni_JAP_1998} and thus been absorbed into $k_{1,2}$ as $k_1=k_1^0+(D_y-D_z)$ and $k_2=k_2^0+(D_x-D_y)$\cite{jlu_EPL_2009,jlu_JAP_2010,jlu_PRB_2016}, where $k_{1,2}^0$ are the magnetic crystalline anisotropy coefficients. The external field $\mathbf{H}_{\mathrm{ext}}$ has two components: the axial driving field $\mathbf{H}_{\parallel}\equiv H_1 \mathbf{e}_z$ and the TMF with general form

\begin{equation}\label{TMF_general}
\mathbf{H}_{\perp}=H_{\perp}(z,t)\left[\cos\Phi(z,t)\mathbf{e}_x+\sin\Phi(z,t)\mathbf{e}_y\right].
\end{equation}
The time evolution of $\mathbf{M}(\mathbf{r},t)$ is described by the Landau-Lifshitz-Gilbert (LLG) euqation\cite{Gilbert_IEEE_2004} as

\begin{equation}\label{LLG_vector}
\frac{\partial \mathbf{m}}{\partial t}=-\gamma \mathbf{m} \times \mathbf{H}_{\mathrm{eff}}+\alpha \mathbf{m}\times \frac{\partial \mathbf{m}}{\partial t},
\end{equation}
where $\alpha$ phenomenologically describes magnetic damping strength, $\gamma>0$ is the absolute value of electron's gyromagnetic ratio and $\mathbf{H}_{\mathrm{eff}}=-\left(\delta\mathcal{E}_{\mathrm{tot}}/\delta\mathbf{M}\right)/\mu_0$ is the effective field.

When system temperature is far below Curie point, the saturation magnetization $M_s$ of magnetic materials can be viewed as constant. Thus $\mathbf{M}(\mathbf{r},t)$ is fully described by its polar angle $\theta(\mathbf{r},t)$ and azimuthal angle $\phi(\mathbf{r},t)$. In addition, for thin enough nanostrips (where TDWs dominate) the inhomogeneity in cross section can be ignored
thus make them become quasi one-dimensional (1D) systems ($\mathbf{r}\rightarrow z$). Then reasonably one has $(\nabla\mathbf{m})^2\equiv(\nabla_z\mathbf{m})^2=(\theta')^2+\sin^2\theta(\phi')^2$ in which a prime means spatial derivative to $z$. After the transition from the global Cartesian coordinate system ($\mathbf{e}_x,\mathbf{e}_y,\mathbf{e}_z$) to the local spherical coordinate system ($\mathbf{e}_{\mathbf{m}},\mathbf{e}_{\theta},\mathbf{e}_{\phi}$), the effective field $\mathbf{H}_{\mathrm{eff}}$ reads

\begin{subequations}\label{H_eff}
\begin{align}
\mathbf{H}_{\mathrm{eff}}&=H_{\mathrm{eff}}^{\mathbf{m}}\mathbf{e}_{\mathbf{m}}+H_{\mathrm{eff}}^{\theta}\mathbf{e}_{\theta}+H_{\mathrm{eff}}^{\phi}\mathbf{e}_{\phi},     \\
H_{\mathrm{eff}}^{\mathbf{m}}&=H_1\cos\theta+H_{\perp}(z,t)\sin\theta\cos\left[\Phi_{\perp}(z,t)-\phi\right]+k_1 M_s-M_s\sin^2\theta\left(k_1+k_2\cos^2\phi\right)   \nonumber  \\
& \quad    \quad    -\frac{2J}{\mu_0 M_s}(\theta'^2+\sin^2\theta\phi'^2)^2,       \\
H_{\mathrm{eff}}^{\theta}&=-H_1\sin\theta+H_{\perp}(z,t)\cos\theta\cos\left[\Phi_{\perp}(z,t)-\phi\right]-M_s\sin\theta\cos\theta\left(k_1+k_2\cos^2\phi\right)   \nonumber  \\
& \quad    \quad    +\frac{2J}{\mu_0 M_s}(\theta''-\sin\theta\cos\theta\phi'^2)\equiv -\mathcal{B},       \\
H_{\mathrm{eff}}^{\phi}&=H_{\perp}(z,t)\sin\left[\Phi_{\perp}(z,t)-\phi\right]+k_2 M_s\sin\theta\sin\phi\cos\phi+\frac{2J}{\mu_0 M_s}\frac{1}{\sin\theta}\left(\sin^2\theta\cdot\phi'\right)'\equiv \mathcal{A}.
	\end{align}
\end{subequations}
Put it back into Eq. (\ref{LLG_vector}), the vectorial LLG equation turns to its scalar counterparts,

\begin{subequations}\label{LLG_scalar_v1}
	\begin{align}
	(1+\alpha^2)\dot{\theta}/\gamma         &=\mathcal{A}-\alpha\mathcal{B}, \\
	(1+\alpha^2)\sin\theta\dot{\phi}/\gamma &=\mathcal{B}+\alpha\mathcal{A},
	\end{align}
\end{subequations}
or equivalently

\begin{subequations}\label{LLG_scalar_v2}
	\begin{align}
	\dot{\theta}+\alpha\sin\theta\dot{\phi} &=\gamma\mathcal{A}, \\
	\sin\theta\dot{\phi}-\alpha\dot{\theta} &=\gamma\mathcal{B},
	\end{align}
\end{subequations}
where a dot means time derivative. These equations are all what we need for our work is this paper.

\section{Results}

In this section, we present in details how to engineer pTDWs with arbitrary tilting attitude by properly tailoring TMF profile along strip axis. As mentioned in Section \ref{introduction}, here we fix the TMF orientation (thus $\Phi_{\perp}(z,t)\equiv\Phi_0$) and allow its strength tunable along strip axis, which is much easier to realize in real experiments. Both statics and axial-field-driven dynamics of pTDWs will be systematically investigated.

\subsection{Statics}\label{Statics}
\unskip
From the roadmap of field-driven DW motion in nanostrips\cite{jlu_EPL_2009}, in the absence of axial driving fields a TDW will finally evolve into its static configurations ($\dot{\theta}=\dot{\phi}=0$) under time-independent TMFs ($H_{\perp}(z,t)\equiv H_{\perp}(z)$). For Eq. (\ref{LLG_scalar_v2}) this means $\mathcal{A}=\mathcal{B}=0$.
In the absence of any TMF ($H_{\perp}(z)\equiv 0$), the static TDW is a pTDW lying in easy plane with the well-known Walker's profile\cite{Walker_JAP_1974},

\begin{equation}\label{Static_profile_noTMF}
\theta(z)=2\arctan e^{\eta\frac{z-z_0}{\Delta_0}},\quad \phi(z)\equiv n\pi/2,
\end{equation}
where $\Delta_0\equiv \sqrt{2J/(\mu_0 k_1 M_s^2)}$ is the pTDW width, $z_0$ is the wall center, $\eta=+1(-1)$ denotes head-to-head (tail-to-tail) pTDWs and $n=+1(-1)$ is the wall polarity (sign of $\langle m_y\rangle$). However, if we want to realize a static pTDW with arbitrary tilting attitude, i.e. $\phi(z)\equiv \phi_{\mathrm{d}}$, well-tailored position-dependent TMF profile must be exerted.

\subsubsection{Boundary condition}
As the first step, we need the boundary condition of this pTDW, which means the magnetization orientation in the two domains at both ends of the strip. Without losing generality, our investigations are performed for head-to-head walls and $0<\phi_{\mathrm{d}}<\pi/2$. In the two domains, the orientation of magnetization should be uniform, meaning that the azimuthal angle satisfies $\phi(z)\equiv\phi_{\mathrm{d}}$, while the polar angle in the left (right) domain takes the value of $\theta_{\mathrm{d}}$ ($\pi-\theta_{\mathrm{d}}$). Meantime, the TMF strength should be constant ($H_{\perp}(z)\rightarrow H_{\perp}^{\mathrm{d}}$) in these two domains. Then $\mathcal{A}=\mathcal{B}=0$ becomes

\begin{subequations}\label{Statics_AandBeq0_in2domains}
	\begin{align}
	H_{\perp}^{\mathrm{d}}\sin(\phi_{\mathrm{d}}-\Phi_0) &=k_2 M_s\sin\theta_{\mathrm{d}}\sin\phi_{\mathrm{d}}\cos\phi_{\mathrm{d}}, \\
	H_{\perp}^{\mathrm{d}}\cos(\Phi_0-\phi_{\mathrm{d}}) &=M_s\sin\theta_{\mathrm{d}}(k_1+k_2\cos^2\phi_{\mathrm{d}}).
	\end{align}
\end{subequations}
The solution to the above equation set provides the TMF profile in the two domains as

\begin{equation}\label{Static_profile_withTMF_in2domains}
\Phi_0=\arctan\left(\frac{k_1}{k_1+k_2}\cdot\tan\phi_{\mathrm{d}}\right), \quad H_{\perp}^{\mathrm{d}}=H_{\perp}^{\mathrm{max}}\cdot\sin\theta_{\mathrm{d}},
\end{equation}
with

\begin{equation}\label{Static_profile_withTMF_Hperp_max}
H_{\perp}^{\mathrm{max}}=M_s\sqrt{k_1^2\sin^2\phi_{\mathrm{d}}+(k_1+k_2)^2\cos^2\phi_{\mathrm{d}}}.
\end{equation}
Eq. (\ref{Static_profile_withTMF_in2domains}) indicates that in both domains, TMF should be farther away from the easy plane than the magnetization. Meanwhile, the existence condition of the pTDW ($\theta_{\mathrm{d}}\ne\pi/2$) requires that TMF strength in domains has an upper limit,

\begin{equation}\label{Static_profile_withTMF_Hperp.lt.HperpMax}
H_{\perp}^{\mathrm{d}}<H_{\perp}^{\mathrm{max}}.
\end{equation}

\subsubsection{Static pTDW profile}
Note that we have fixed TMF orientation to be $\Phi_0$, therefore in pTDW region $\mathcal{A}=\mathcal{B}=0$ becomes

\begin{subequations}\label{Statics_AandBeq0_inpTDW_v1}
	\begin{align}
	0&=H_{\perp}(z)\sin\left(\Phi_0-\phi\right)+k_2 M_s\sin\theta\sin\phi\cos\phi+\frac{2J}{\mu_0 M_s}\frac{1}{\sin\theta}\left(\sin^2\theta\cdot\phi'\right)',      \\
	\frac{2J}{\mu_0 M_s}\theta''&=-H_{\perp}(z)\cos\theta\cos\left(\Phi_0-\phi\right)+M_s\sin\theta\cos\theta\left(k_1+k_2\cos^2\phi\right)+\frac{2J}{\mu_0 M_s}\sin\theta\cos\theta\phi'^2. 
	\end{align}
\end{subequations}
Since we are considering pTDWs with uniform tilting attitude $\phi(z)\equiv\phi_{\mathrm{d}}$, then the above equations become

\begin{subequations}\label{Statics_AandBeq0_inpTDW_v2}
	\begin{align}
	H_{\perp}(z)\sin\left(\phi_{\mathrm{d}}-\Phi_0\right)&=k_2 M_s\sin\theta\sin\phi_{\mathrm{d}}\cos\phi_{\mathrm{d}},  \\
	\frac{2J}{\mu_0 M_s}\theta''&=-H_{\perp}(z)\cos\theta\cos\left(\Phi_0-\phi_{\mathrm{d}}\right)+M_s\sin\theta\cos\theta\left(k_1+k_2\cos^2\phi_{\mathrm{d}}\right). 
	\end{align}
\end{subequations}
Combing Eqs. (\ref{Statics_AandBeq0_in2domains}a) and (\ref{Statics_AandBeq0_inpTDW_v2}a), one has

\begin{equation}\label{Statics_Hperp_profile_v1}
H_{\perp}(z)=\frac{H_{\perp}^{\mathrm{d}}}{\sin\theta_{\mathrm{d}}}\cdot\sin\theta(z)=H_{\perp}^{\mathrm{max}}\cdot\sin\theta(z).
\end{equation}
Putting it back into Eq. (\ref{Statics_AandBeq0_inpTDW_v2}b) and considering Eq. (\ref{Statics_AandBeq0_in2domains}b), it turns out that

\begin{equation}\label{Statics_theta_2ndDerivative}
\frac{2J}{\mu_0 M_s}\theta''=\frac{\sin\theta\cos\theta}{\sin\theta_{\mathrm{d}}}\left[M_s\sin\theta_{\mathrm{d}}(k_1+k_2\cos^2\phi_{\mathrm{d}})-H_{\perp}^{\mathrm{d}}\cos(\Phi_0-\phi_{\mathrm{d}})\right]=0,
\end{equation}
which means $\theta(z)$ is linear in pTDW region,

\begin{equation}\label{Statics_theta}
\theta(z)=C_1+C_2\cdot(z-z_0),
\end{equation}
where $z_0$ is the pTDW center.

\begin{figure}[H]
	\centering
	\includegraphics[width=10 cm]{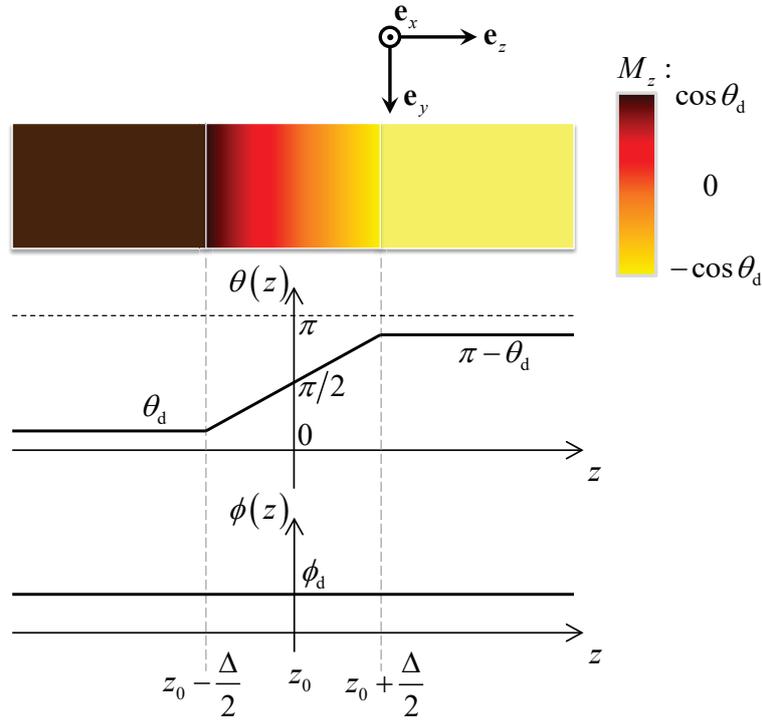}
	\caption{Illustration of pTDW profile with arbitrary titling attitude $\phi_{\mathrm{d}}$, controllable width $\Delta$ and linear polar angle distribution from $\theta_{\mathrm{d}}$ to $\pi-\theta_{\mathrm{d}}$. The color chart indicates the variation of $M_z$ component along strip axis from $\cos\theta_{\mathrm{d}}$ in the left domain to $-\cos\theta_{\mathrm{d}}$ in the right domain.}
\end{figure}   

It is worth noting that in nearly all existing literatures, the boundary between ``domains" and ``domain walls" in nanostrips is not clear (or abrupt) since $\theta(z)$ and $\phi(z)$ and their derivatives are all continuous there. However, Eqs. (\ref{Statics_Hperp_profile_v1}) to (\ref{Statics_theta}) provide us an opportunity to realize a pTDW with clear boundary and tunable width, as depicted in Figure. 2. In summary, under the following TMF distribution

\begin{equation}\label{Statics_Hperp_profile_v2}
H_{\perp}(z)=
\left\{
\begin{array}{cc}
H_{\perp}^{\mathrm{d}},             &    z<z_0-\frac{\Delta}{2} \\
H_{\perp}^{\mathrm{max}}\cdot\sin\left\{\theta_{\mathrm{d}}+\frac{\pi-2\theta_{\mathrm{d}}}{\Delta}\left[z-\left(z_0-\frac{\Delta}{2}\right)\right]\right\},     &    z_0-\frac{\Delta}{2}<z<z_0+\frac{\Delta}{2} \\
H_{\perp}^{\mathrm{d}},             &    z>z_0+\frac{\Delta}{2} \\
\end{array}
\right., \quad \Phi_{\perp}(z)\equiv\Phi_0,
\end{equation}
a pTDW with the following profile will emerge in the nanostrip,

\begin{equation}\label{Statics_pTDW_profile}
\theta_0(z)=
\left\{
\begin{array}{cc}
\theta_{\mathrm{d}},             &    z<z_0-\frac{\Delta}{2} \\
\theta_{\mathrm{d}}+\frac{\pi-2\theta_{\mathrm{d}}}{\Delta}\left[z-\left(z_0-\frac{\Delta}{2}\right)\right],     &    z_0-\frac{\Delta}{2}<z<z_0+\frac{\Delta}{2} \\
\pi-\theta_{\mathrm{d}},             &    z>z_0+\frac{\Delta}{2} \\
\end{array}
\right., \quad \phi_0(z)\equiv\phi_{\mathrm{d}}.
\end{equation}

Interestingly, the above pTDW has the following features: (i) an arbitrary tilting attitude $\phi_{\mathrm{d}}$. (ii) a fully controllable width $\Delta$ and (iii) two clear boundaries ($z_0\pm\Delta/2$) with the two adjacent domains. Note that the magnetization and TMF at $z_0\pm\Delta/2$ are both continuous, but $\nabla_z\mathbf{m}$ is not. This inevitably leads to a finite jump of exchange energy density right there. 

However, the pTDW has a critical width $\Delta_{\mathrm{c}}$ under which the entire strip has lower magnetic energy compared with the single-domain state under the UTMF with strength $H_{\perp}^{\mathrm{d}}$ and orientation $\Phi_0$. To see this, we integrate $\mathcal{E}_{\mathrm{tot}}^{\mathrm{pTDW}}-\mathcal{E}_{\mathrm{tot}}^{\mathrm{domain}}$ over the entire strip and thus

\begin{equation}\label{E_difference}
\Delta E=\frac{k_1 \mu_0 M_s^2}{2}\cdot\left[\left(\Delta_0\right)^2(\pi-2\theta_{\mathrm{d}})^2\frac{1}{\Delta}-\frac{\sin 2\theta_{\mathrm{d}}+(\pi-2\theta_{\mathrm{d}})\cos 2\theta_{\mathrm{d}}}{2(\pi-2\theta_{\mathrm{d}})}\left(1+\frac{k_2}{k_1}\cos^2\phi_{\mathrm{d}}\right)\Delta\right].
\end{equation}
Obviously, there exists a critical pTDW width

\begin{equation}\label{Delta_c}
\Delta_{\mathrm{c}}\equiv\Delta_0\cdot\left(1+\frac{k_2}{k_1}\cos^2\phi_{\mathrm{d}}\right)^{-\frac{1}{2}}\cdot \kappa(\theta_{\mathrm{d}}),\quad \kappa(\theta_{\mathrm{d}}) \equiv \sqrt{\frac{2(\pi-2\theta_{\mathrm{d}})^3}{\sin 2\theta_{\mathrm{d}}+(\pi-2\theta_{\mathrm{d}})\cos 2\theta_{\mathrm{d}}}}.
\end{equation}
As $H_{\perp}^{\mathrm{d}}\rightarrow H_{\perp}^{\mathrm{max}}$, by defining $\frac{H_{\perp}^{\mathrm{d}}}{H_{\perp}^{\mathrm{max}}}=1-\epsilon$ we have $\theta_{\mathrm{d}}=\arcsin\frac{H_{\perp}^{\mathrm{d}}}{H_{\perp}^{\mathrm{max}}}\approx\frac{\pi}{2}-\sqrt{2\epsilon}$, thus $\sin 2\theta_{\mathrm{d}}\approx 2\sqrt{2\epsilon}$, $\cos 2\theta_{\mathrm{d}}\approx -1+4\epsilon$ and $\pi-2\theta_{\mathrm{d}}\approx 2\sqrt{2\epsilon}$. Putting all these approximations back into $\kappa$ in Eq. (\ref{Delta_c}), we finally get $\kappa\rightarrow 2$ which leads to a finite critical pTDW $\Delta_{\mathrm{c}}$. As a result, we can always make the pTDW energetically preferred by setting $\Delta>\Delta_{\mathrm{c}}$ (thus $\Delta E<0$).

\subsubsection{Stability analysis}\label{Stability_analysis}
To make the explorations on statics complete and self-consistent, we need to perform stability analysis on the pTDW profile in Eq. (\ref{Statics_pTDW_profile}). For simplicity, the variations on $\theta(z)$ and $\phi(z)$ are processed separately. In the first step, $\phi(z)\equiv\phi_0$ is fixed (thus $\dot{\phi}\equiv 0$) and suppose the polar angle departs from its static profile as

\begin{equation}\label{Stability_statics_theta_0}
\theta=\theta_0+\delta\theta.
\end{equation}
Putting it back into Eq. (\ref{LLG_scalar_v2}b), by noting that $\dot{\phi}_0=0$ and $\dot{\theta}_0=0$, one has

\begin{equation}\label{Stability_statics_theta_1}
\sin\theta\dot{\phi}-\alpha\dot{\theta} =\gamma\mathcal{B} \Rightarrow \frac{\alpha}{\gamma}\frac{\partial(\delta\theta)}{\partial t}=-\mathcal{B}.
\end{equation}
On the other hand, in pTDW region $\theta_0$ satisfies Eq. (\ref{Statics_AandBeq0_inpTDW_v2}b). After performing series expansion on $\mathcal{B}$ around $\theta_0$ and preserving up to linear terms of $\delta\theta$, we finally get

\begin{equation}\label{Stability_statics_theta_2}
\frac{\alpha}{\gamma}\frac{\partial(\delta\theta)}{\partial t}\approx \left[-M_s\cos^2\theta_0(k_1+k_2\cos^2\phi_0)+\frac{2J}{\mu_0 M_s}\frac{(\delta\theta)''}{\delta\theta} \right]  \cdot\delta\theta.
\end{equation}
Obviously, when

\begin{equation}\label{Stability_statics_theta_3}
\left|\frac{(\delta\theta)''}{\delta\theta}\right|<\frac{\cos^2\theta_0(1+k_2\cos^2\phi_0/k_1)}{(\Delta_0)^2},
\end{equation}
$\delta\theta$ fades out as times goes by. This implies that when the variation $\delta\theta$ is not too abrupt, $\theta_0$ is stable. In fact, most variations satisfy this demand.  For example, both tiny global translations along $z-$axis and slight local variations proportional to $z-z_0$ make $(\delta\theta)''\equiv 0$ thus assure the stability around $\theta_0$.
	
In the second step, we keep $\theta(z)\equiv\theta_0$ and let the azimuthal angle varies as follows

\begin{equation}\label{Stability_statics_phi_0}
\phi=\phi_0+\delta\phi.
\end{equation}
Substituting it into Eq. (\ref{LLG_scalar_v2}a), by recalling that $\dot{\theta}_0=0$ and $\dot{\phi}_0=0$, we have

\begin{equation}\label{Stability_statics_phi_1}
\dot{\theta}+\alpha\sin\theta\dot{\phi} =\gamma\mathcal{A} \Rightarrow \frac{\alpha}{\gamma}\frac{\partial(\delta\phi)}{\partial t}=\frac{\mathcal{A}}{\sin\theta_0}.
\end{equation}
Remember in pTDW region $\theta_0$ and $\phi_0$ satisfy Eq. (\ref{Statics_AandBeq0_inpTDW_v2}a). By performing series expansion on $\mathcal{A}$ about $\phi_0$ and at most keeping linear terms of $\delta\theta$, one has

\begin{equation}\label{Stability_statics_phi_2}
\frac{\alpha}{\gamma}\frac{\partial(\delta\phi)}{\partial t}\approx \left[-M_s(k_1+k_2\sin^2\phi_0)+\frac{2J}{\mu_0 M_s}\frac{2\cot\theta_0\cdot\theta'_0\cdot(\delta\phi)'+(\delta\phi)''}{\delta\phi} \right]  \cdot\delta\phi.
\end{equation}
Similarly, if $\delta\phi$ does not varies too abruptly, that is

\begin{equation}\label{Stability_statics_phi_3}
\left|\frac{2\cot\theta_0\cdot\theta'_0\cdot(\delta\phi)'+(\delta\phi)''}{\delta\phi}\right|<\frac{1+k_2\sin^2\phi_0/k_1}{(\Delta_0)^2},
\end{equation}
the pTDW is stable around $\phi_0$, which confirms the feasibility of engineering pTDWs in magnetic nanostrips. In particular, tiny global rotations around $z-$axis or slight local twistings proportional to $z-z_0$ will not drive pTDW away from its static profile shown in Eq. (\ref{Statics_pTDW_profile}).

\subsubsection{Numerical confirmations}
To confirm the above theoretical analysis, we perform numerical simulations using the OOMMF micromagnetics package\cite{OOMMF}. In our simulations, the nanostrip is 5 nm thick, 100 nm wide and 1 $\mu$m long, which is quite common in real experiments. The three average demagnetization factors are: $D_x=0.00661366$, $D_y=0.07002950$ and $D_z=0.92335684$\cite{Aharoni_JAP_1998}. Magnetic parameters are as follows: $M_s=500$ kA/m, $J=40\times 10^{-12}$ J/m, $K_1=\mu_0 k_1^0 M_s^2/2=200$ kJ/m$^3$, $K_2=\mu_0 k_2^0 M_s^2/2=50$ kJ/m$^3$ and $\alpha=0.1$ to speed up the simulation. Throughout the entire calculation, the strip is discretized into $5\times5\times5$ nm$^3$ cells and all magnetic intensive quantities evaluated at each cell are the average of their continuous counterparts over the cell volume. In all figures, $z_0$ denotes the wall center which is the algebraic average of the central positions ($\phi(z)=\pi/2$) of each layer (row of cells with a certain $y$-coordinate). At last, the external TMF at each cell is the value from Eq. (\ref{Statics_Hperp_profile_v2}) at the cell center.

We aim to realize a pTDW with tilting attitude $\phi_{\mathrm{d}}\equiv\pi/4$ and boundary condition $\theta_{\mathrm{d}}\equiv\pi/6$ under the TMF profile in Eq. (\ref{Statics_Hperp_profile_v2}). To do this, firstly simple algebra provides us $\Delta_0=13.80$ nm (14.14 nm) when the demagnetization is (not) considered. Then the critical pTDW width $\Delta_{\mathrm{c}}=35.66$ nm (41.31 nm) for each case. Therefore we set the pTDW width as $\Delta=100$ nm to assure energetic preference. We have performed simulations for both cases in which magnetostatic effect is included or not. At each case, a standard head-to-head N\'{e}el wall with width 20 nm is generated at the strip center beforehand. After it relaxes to its stable profile, a time-independent TMF described by Eq. (\ref{Statics_Hperp_profile_v2}) is exerted onto each calculation cell of this strip. The magnetization texture then begin to evolve accompanied by the decreasing total magnetic energy due to the Gilbert damping process. We set the convergence strategy as $|\mathbf{m}\times\mathbf{H}_{\mathrm{tot}}|/M_s<10^{-7}$, which is accurate enough. The results are plotted in Figure 3(a) and 3(b), respectively.

In the simpler case, the pTDW profile under TMF distribution described in Eq. (\ref{Statics_Hperp_profile_v2}) with $\phi_{\mathrm{d}}\equiv\pi/4$, $\theta_{\mathrm{d}}\equiv\pi/6$ and $\Delta=100$ nm in the absence of demagnetization is plotted in Figure 3(a). The solid black and red lines are the analytical polar and azimuthal distributions from Eq. (\ref{Statics_pTDW_profile}), respectively. The open circles are numerical data from OOMMF simulation. Clearly the planar nature of wall is reproduced very well. For polar angle, the linear behavior near pTDW center is unambiguous. While the discontinuity in polar angle derivative at pTDW border ($z_0\pm$ 50 nm) is weakened due to the inevitable ``discretized sampling" of TMF at calculation cells during numerical simulations. In summary one may clearly see that the numerics and analytics fit very well.

\begin{figure}[H]
	\centering
	\includegraphics[width=10 cm]{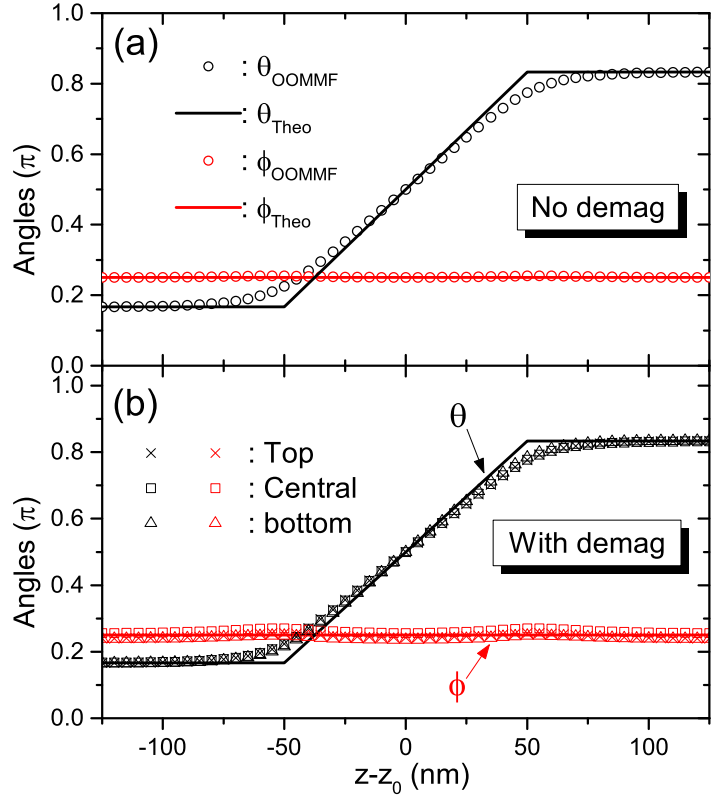}
	\caption{Comparisons between analytical (solid lines) and numerical (hollow symbols) pTDW profiles under TMF in Eq. (\ref{Statics_Hperp_profile_v2}) with $\phi_{\mathrm{d}}\equiv\pi/4$, $\theta_{\mathrm{d}}\equiv\pi/6$ and $\Delta=100$ nm: (a) without demagnetization, (b) with demagnetization. The magnetic parameters are as follows: $M_s=500$ kA/m, $J=40\times 10^{-12}$ J/m, $K_1=\mu_0 k_1^0 M_s^2/2=200$ kJ/m$^3$, $K_2=\mu_0 k_2^0 M_s^2/2=50$ kJ/m$^3$ and $\alpha=0.1$.}
\end{figure}  

Then we switch on the magnetostatic interaction (demagnetization). Due to the complicated dipole-dipole interaction, the magnetization orientation in the strip cross section differs a little (not too much since the strip is thin enough). We then calculate the polar and azimuthal angles for three typical layer (rows of cells with the same $y-$coordinate): top, central and bottom. The resulting data are depicted in Figure 3(b) by different discrete hollow symbols: crosses, squares and triangles. It turns out that they overlap each other nicely and match the analytical profiles quite well. This not only reproves the validity of TMF in Eq. (\ref{Statics_Hperp_profile_v2}) for realizing pTDW in Eq. (\ref{Statics_pTDW_profile}) under more complex situations, but also shows once again the feasibility of simplifying magnetostatic energy by local quadratic terms in thin enough nanostrips.

\subsection{Axial-field-driven dynamics}\label{Dynamics}
From the roadmap of field-driven DW dynamics\cite{jlu_EPL_2009}, an axial magnetic field is crucial for driving pTDWs to move along strip axis thus realizing bit-switchings in magnetic nanodevices based on them. We focus on the traveling-wave mode of pTDWs in which their profile is generalized directly from Eq. (\ref{Statics_pTDW_profile}) by allowing $z_0$ to depend on time meantime leaving the rest unchanged. To preserve the pTDW profile, the TMF distribution is suggested to take the same form as in Eq. (\ref{Statics_Hperp_profile_v2}) but with the generalized $z_0$, which means that TMF moves along with the pTDW sharing the same velocity. In this section, the dynamics of these pTDWs are systematically investigated under two strategies: 1D collective coordinate model (1D-CCM)\cite{Tatara_JPDAP_2011} and 1D asymptotic expansion method (1D-AEM)\cite{jlu_PRB_2016,limei_srep_2017,Goussev_PRB_2013,Goussev_Royal_2013}. As will be shown below, they provide the same result which confirms the feasibility of both approaches.

\subsubsection{1D-CCM}\label{1D-CCM}
Historically, 1D-CCM plays important role in the exploration of TDW dynamics for both field-driven and current-driven cases. Generally it treats the center, tilting attitude and with of a DW as independent collective variables of the system Lagrangian or the resulting dynamical equations (i.e. LLG equation). The classical Walker ansatz (which is indeed a pTDW profile) in the absence of any TMFs is the first example and turns out to be the rigorous solution of LLG equation. In the presence of UTMFs, generally no rigorous solutions exist due to the mismatch between symmetries in different energy terms. In most theoretical works, pTDWs with quasi-Walker profiles are often proposed to mimic the real complicated magnetization distribution. However, in Section \ref{Statics} it has been shown that the Walker ansatz is not the only choice that a pTDW can preceed. In this subsection, we provide the pTDW velocity with comoving TMF profile in the framework of 1D-CCM.

Before the main context, we want to point out that to preserve the planar feature of these walls, the strength of axial driving field should not be too high. To see this, we revisit the boundary condition in the two domains in the presence of axial driving field $H_1$. Note that although in pTDW region, $\mathbf{H}_{\mathrm{eff}}$ is not parallel with $\mathbf{m}$ (otherwise the wall will not move), however in both domains it holds since magnetization does not vary with time, hence $\mathcal{A}=\mathcal{B}=0$ therein. After redefining the polar and azimuthal angles of magnetization in the left domain as $\tilde{\theta}_{\mathrm{d}}$ and $\tilde{\phi}_{\mathrm{d}}$ ($\pi-\tilde{\theta}_{\mathrm{d}}$ and $\tilde{\phi}_{\mathrm{d}}$ in the right domain), one has

\begin{subequations}\label{Dynamics_AandBeq0_in2domains}
	\begin{align}
	0&=H_{\perp}^{\mathrm{d}}\sin(\Phi_0-\tilde{\phi}_{\mathrm{d}})+k_2 M_s\sin\tilde{\theta}_{\mathrm{d}}\sin\tilde{\phi}_{\mathrm{d}}\cos\tilde{\phi}_{\mathrm{d}}, \\
	0&=H_1\sin\tilde{\theta}_{\mathrm{d}}-H_{\perp}^{\mathrm{d}}\cos\tilde{\theta}_{\mathrm{d}}\cos(\Phi_0-\tilde{\phi}_{\mathrm{d}})+M_s\sin\tilde{\theta}_{\mathrm{d}}\cos\tilde{\theta}_{\mathrm{d}}(k_1+k_2\cos^2\tilde{\phi}_{\mathrm{d}}),
	\end{align}
\end{subequations}
Obviously, only when $H_1\ll \min[H_{\perp}^{\mathrm{d}},M_s]$ one has $\tilde{\theta}_{\mathrm{d}}\approx\theta_{\mathrm{d}}$ and $\tilde{\phi}_{\mathrm{d}}\approx\phi_{\mathrm{d}}$. Then after the generalization of collective coordinate $z_0$ from constant to time-dependent, the pTDW in Eq. (\ref{Statics_pTDW_profile}) is expected to move along strip axis under the comoving TMF in Eq. (\ref{Statics_Hperp_profile_v2}) with the velocity equal to $\mathrm{d}z_0/\mathrm{d}t$. 

To determine wall velocity in traveling-wave mode, we perform time derivative of the pTDW profile which gives

\begin{equation}\label{Dynamics_pTDW_profile_time_derivative}
\dot{\theta}(z,t)=
\left\{
\begin{array}{cc}
0,             &    z<z_0-\frac{\Delta}{2} \\
-\frac{\pi-2\theta_{\mathrm{d}}}{\Delta}\cdot\frac{\mathrm{d}z_0}{\mathrm{d}t},     &    z_0-\frac{\Delta}{2}<z<z_0+\frac{\Delta}{2} \\
0,             &    z>z_0+\frac{\Delta}{2} \\
\end{array}
\right., \quad \dot{\phi}(z,t)\equiv 0.
\end{equation}
From Eq. (\ref{LLG_scalar_v1}b), the traveling-mode condition $\dot{\phi}(z,t)\equiv 0$ leads to $\mathcal{A}=-\mathcal{B}/\alpha$. Putting back into Eq. (\ref{LLG_scalar_v1}a), it turns out that $-\alpha\dot{\theta}(z,t)/\gamma=\mathcal{B}$. Substituting Eq. (\ref{Dynamics_pTDW_profile_time_derivative}) into it, one has

\begin{equation}\label{Dynamics_velocity_v1}
\frac{\alpha}{\gamma}\cdot\frac{\pi-2\theta_{\mathrm{d}}}{\Delta}\cdot\frac{\mathrm{d}z_0}{\mathrm{d}t}= H_1\sin\theta-H_{\perp}(z,t)\cos\theta\cos\left(\Phi_0-\phi\right)+M_s\sin\theta\cos\theta\left(k_1+k_2\cos^2\phi\right)-\frac{2J}{\mu_0 M_s}\theta''.
\end{equation}
Note that the generalized TMF configuration and the resulting pTDW profile still satisfy Eq. (\ref{Statics_AandBeq0_inpTDW_v2}b), thus eliminate the last three terms in the right hand side of the above equation. Then after integrating Eq. (\ref{Dynamics_velocity_v1}) over the pTDW region, $z\in\left(z_0-\frac{\Delta}{2},z_0+\frac{\Delta}{2}\right)$, and noting that $\int_{z_0-\Delta/2}^{z_0+\Delta/2}1\mathrm{d}z=\Delta$, $\int_{z_0-\Delta/2}^{z_0+\Delta/2}\sin\theta\mathrm{d}z=2\Delta\cos\theta_{\mathrm{d}}/(\pi-2\theta_{\mathrm{d}})$, we finally get

\begin{equation}\label{Dynamics_velocity_v2}
V_{\mathrm{a}}\equiv\frac{\mathrm{d}z_0}{\mathrm{d}t}=\frac{\gamma\Delta}{\alpha}\cdot\omega(\theta_{\mathrm{d}})\cdot H_1,\quad \omega(\theta_{\mathrm{d}})\equiv\frac{2\cos\theta_{\mathrm{d}}}{(\pi-2\theta_{\mathrm{d}})^2}.
\end{equation}
Next we examine the asymptotic behavior of the boosting factor $\omega(\theta_{\mathrm{d}})$ when $H_{\perp}^{\mathrm{d}}\rightarrow H_{\perp}^{\mathrm{max}}$. Suppose again $\frac{H_{\perp}^{\mathrm{d}}}{H_{\perp}^{\mathrm{max}}}=1-\epsilon$, then $\cos\theta_{\mathrm{d}}\approx\sqrt{2\epsilon}$ and $\pi-2\theta_{\mathrm{d}}\approx 2\sqrt{2\epsilon}$. Putting them back into Eq. (\ref{Dynamics_velocity_v2}), we finally have

\begin{equation}\label{Dynamics_boosting_factor}
\omega(\theta_{\mathrm{d}})\approx\frac{1}{2\sqrt{2\epsilon}}\rightarrow +\infty,
\end{equation}
as $\epsilon\rightarrow 0^+$. This confirms the boosting effect of these TMFs on axial propagation of pTDWs.

At last, stability analysis to dynamical pTDW profile under comoving TMFs takes the same format as static case and thus has been omitted for saving space. It turns out that for profile variations which are not too abrupt, the traveling-wave mode of pTDW is also stable. This is really important for potential commercial applications of these pTDWs.

\subsubsection{1D-AEM}\label{1D-AEM}
Next we recalculate the pTDW velocity in traveling-wave mode with the help of 1D-AEM. In this approach, the dynamical behavior of pTDWs is viewed as the response of their static profiles to external stimuli. Therefore it is the manifestation of linear response framework in nanomagnetism and should be suitable for exploring traveling-wave mode of pTDWs under small axial driving fields. Note that the TMF distribution in Eq. (\ref{Statics_Hperp_profile_v2}) indicates that at the pTDW center TMF strength reaches $H_{\perp}^{\mathrm{max}}$ which is finite, thus we rescale the axial driving field and pTDW axial velocity simultaneously,

\begin{equation}\label{1D_AEM_scaling}
H_1=\epsilon h_1,\quad V_{\mathrm{b}}=\epsilon v_{\mathrm{b}},
\end{equation}
in which $\epsilon$ is a dimensionless infinitesimal. This means a slight external stimulus ($H_1$) will lead to a weak response of the system, that is, a slow velocity ($V_{\mathrm{b}}$) of pTDW axial motion. We concentrate on traveling-wave mode of pTDWs thus define the traveling coordinate

\begin{equation}\label{1D_AEM_finiteTMF_xi}
\xi\equiv z-V_{\mathrm{b}} t=z-\epsilon v_{\mathrm{b}} t.
\end{equation}
Meantime the TMF distribution takes the same one as in Eq. (\ref{Statics_Hperp_profile_v2}), except for the generalization of $z\rightarrow \xi$. As a result, the real solution of pTDW can be expanded as follows,

\begin{equation}\label{1D_AEM_series_expansion}
\chi(z,t) = \chi_0(\xi)+\epsilon\chi_1(\xi)+O(\epsilon^2),\quad \chi=\theta(\phi),
\end{equation}
where $\theta_0(\phi_0)$ denote the zeroth-order solutions and should be the static pTDW profile (will see later), while $\theta_1$ and $\phi_1$ are the coefficients of first-order corrections to zeroth-order solutions when $H_1$ is present. Putting them into the LLG equation (\ref{LLG_scalar_v2}) and noting that $\partial \chi/\partial t=(-\epsilon v_{\mathrm{b}})\cdot\partial \chi/\partial \xi$, we have

\begin{subequations}\label{1D_AEM_LLG_scalar_expansion}
	\begin{align}
	(-\epsilon v_{\mathrm{b}})\cdot\left(\frac{\partial \theta_0}{\partial \xi}+\alpha\sin\theta_0\frac{\partial \phi_0}{\partial \xi}\right)+ O(\epsilon^2) &=\gamma \mathcal{A}_0 +\gamma \mathcal{A}_1\cdot\epsilon + O(\epsilon^2), \\
	(-\epsilon v_{\mathrm{b}})\cdot\left(\sin\theta_0\frac{\partial \phi_0}{\partial \xi}-\alpha\frac{\partial \theta_0}{\partial \xi}\right)+ O(\epsilon^2) &=\gamma \mathcal{B}_0 +\gamma \mathcal{B}_1\cdot\epsilon + O(\epsilon^2),
	\end{align}
\end{subequations}
 with 

\begin{subequations}\label{1D_AEM_LLG_scalar_expansion_A0B0}
	\begin{align}
	A_0&=H_{\perp}(\xi)\sin(\Phi_0-\phi_0)+k_2 M_s \sin\theta_0\sin\phi_0\cos\phi_0+\frac{2J}{\mu_0 M_s}\left(2\cos\theta_0\frac{\partial\theta_0}{\partial\xi}\frac{\partial\phi_0}{\partial\xi}+\sin\theta_0\frac{\partial^2\phi_0}{\partial\xi^2}\right), \\
	B_0&=-H_{\perp}(\xi)\cos\theta_0\cos(\Phi_0-\phi_0)-\frac{2J}{\mu_0 M_s}\frac{\partial^2\theta_0}{\partial\xi^2}+k_1 M_s \sin\theta_0\cos\theta_0\left[1+\frac{k_2}{k_1}\cos^2\phi_0+\Delta_0^2\left(\frac{\partial\phi_0}{\partial\xi}\right)^2\right],
	\end{align}
\end{subequations} 
and

\begin{eqnarray}\label{1D_AEM_LLG_scalar_expansion_A1}
	A_1&=&\mathbf{P}\theta_1+\mathbf{Q}\phi_1,  \nonumber  \\
	\mathbf{P}&=&k_2 M_s \cos\theta_0\sin\phi_0\cos\phi_0+\frac{2J}{\mu_0 M_s}\left[2\frac{\partial\phi_0}{\partial\xi}\left(\cos\theta_0\frac{\partial}{\partial\xi}-\sin\theta_0\frac{\partial\theta_0}{\partial\xi}\right)+ \cos\theta_0\frac{\partial^2\phi_0}{\partial\xi^2}\right],  \nonumber \\
	\mathbf{Q}&=&-H_{\perp}(\xi)\cos(\Phi_0-\phi_0)+k_2 M_s\sin\theta_0\cos2\phi_0+\frac{2J}{\mu_0 M_s}\left(2\cos\theta_0\frac{\partial\theta_0}{\partial\xi}\frac{\partial}{\partial\xi}+\sin\theta_0\frac{\partial^2}{\partial\xi^2}\right),
\end{eqnarray}
as well as

\begin{eqnarray}\label{1D_AEM_LLG_scalar_expansion_B1}
B_1&=&h_1\sin\theta_0+\mathbf{R}\theta_1+\mathbf{S}\phi_1,  \nonumber  \\
\mathbf{R}&=&H_{\perp}(\xi)\sin\theta_0\cos(\Phi_0-\phi_0)-\frac{2J}{\mu_0 M_s}\frac{\partial^2}{\partial\xi^2}+k_1 M_s \cos2\theta_0\left[1+\frac{k_2}{k_1}\cos^2\phi_0+\Delta_0^2\left(\frac{\partial\phi_0}{\partial\xi}\right)^2\right],  \nonumber \\
\mathbf{S}&=&-H_{\perp}(\xi)\cos\theta_0\sin(\Phi_0-\phi_0)+k_1 M_s \sin 2\theta_0\left(\Delta_0^2\frac{\partial\phi_0}{\partial\xi}\frac{\partial}{\partial\xi}-\frac{k_2}{k_1}\sin\phi_0\cos\phi_0\right).
\end{eqnarray}

At the zeroth order of $\epsilon$, Eq. (\ref{1D_AEM_LLG_scalar_expansion}) provides $\mathcal{A}_0=\mathcal{B}_0=0$. Combing with the definitions in Eq. (\ref{1D_AEM_LLG_scalar_expansion_A0B0}), its solution is just the pTDW profile in Eq. (\ref{Statics_pTDW_profile}) except for the substitution of $z\rightarrow \xi$. This is not surprising since zeroth-order solution describes the response of system under ``zero" stimulus which is just the static case.

However to obtain the pTDW velocity, we need to proceed to the first order of $\epsilon$. In particular, we have to deal with $\mathbf{R}$ and $\mathbf{S}$ to get the dependence of velocity ($v_{\mathrm{b}}$) on axial driving field ($h_1$). By partially differentiating $\mathcal{B}_0=0$ with respect to $\phi_0$, $\mathbf{S}$ can be simplified to 

\begin{equation}\label{1D_AEM_S_simplified}
\mathbf{S}=\Delta_0^2 k_1 M_s \sin 2\theta_0\left(\frac{\partial\phi_0}{\partial\xi}\frac{\partial}{\partial\xi}-\frac{\partial^2\phi_0}{\partial\xi^2}\right)\equiv 0
\end{equation}
due to the planar nature of walls. On the other hand, the partial derivative of $\mathcal{B}_0=0$ with respect to $\theta_0$ helps to simplify $\mathbf{R}$ to

\begin{equation}\label{1D_AEM_R_simplified_to_L}
\mathbf{R}=\frac{2J}{\mu_0 M_s}\left[-\frac{\partial^2}{\partial\xi^2}+\left(\frac{\partial\theta_0}{\partial\xi}\right)^{-1}\left(\frac{\partial^3\theta_0}{\partial\xi^3}\right)\right]\equiv \mathbf{L},
\end{equation}
which is the 1D self-adjoint Schr\"{o}dinger operator appeared in previous works\cite{jlu_PRB_2016,limei_srep_2017,Goussev_PRB_2013,Goussev_Royal_2013}. Then Eq. (\ref{1D_AEM_LLG_scalar_expansion_B1}) rigorously turns to

\begin{equation}\label{1D_AEM_L_theta1}
\mathbf{L}\theta_1=-h_1\sin\theta_0+(-v_{\mathrm{b}})\cdot\left(-\alpha\frac{\partial \theta_0}{\partial \xi}\right).
\end{equation} 
Again the ``Fredholm alternative" requests the right hand side of the above equation to be orthogonal to the kernel of $\mathbf{L}$ (subspace expanded by $\partial\theta_0/\partial\xi$) for the existence of a solution $\theta_1$, where the inner product in Sobolev space is defined as $\langle f(\xi),g(\xi) \rangle\equiv\int_{\xi=-\infty}^{\xi=+\infty}f(\xi)\cdot g(\xi)\mathrm{d}\xi$. Noting that $\langle\frac{\partial\theta_0}{\partial\xi},\sin\theta_0 \rangle=2\cos\theta_{\mathrm{d}}$ and $\langle\frac{\partial\theta_0}{\partial\xi},\frac{\partial\theta_0}{\partial\xi} \rangle=(\pi-2\theta_{\mathrm{d}})^2/\Delta$, we finally get

\begin{equation}\label{1D_AEM_velocity}
V_{\mathrm{b}}\equiv\frac{\mathrm{d}z_0}{\mathrm{d}t}=\frac{\gamma\Delta}{\alpha}\cdot\frac{2\cos\theta_{\mathrm{d}}}{(\pi-2\theta_{\mathrm{d}})^2}\cdot H_1,
\end{equation}
which is the same as Eq. (\ref{Dynamics_velocity_v2}) from 1D-CCM.

\section{Discussion}
In Section \ref{Dynamics} we point out that under axial driving fields, the pTDW velocity can be considerably increased due to the divergent behavior of the boosting factor $\omega(\theta_{\mathrm{d}})$ when $H_{\perp}\rightarrow H_{\perp}^{\mathrm{max}}$ (see Eq. (\ref{Dynamics_boosting_factor})). Interestingly, the contribution of pTDW width, i.e. $\Delta$, is also an important boosting factor. From Eq. (\ref{Delta_c}) one has a finite critical pTDW width even when $H_{\perp}\rightarrow H_{\perp}^{\mathrm{max}}$. Therefore to further increase the pTDW velocity, broadening the pTDW width should also be effective. 

Second, to realized pTDWs the ``orientation-fixed" strategy proposed here has several advantages comparing with the ``amplitude-fixed" one introduced before\cite{limei_srep_2017}: (i) the wall width can be freely tuned. (ii) the rigorous pTDW profile and the corresponding TMF distribution can be explicitly written out. (iii) the asymptotic behavior of the boosting factor in axial-field-driven case can be analytically explored. (iv) most importantly, the ``orientation-fixed" strategy is much easier to realize in real experiments. 

For example, the following procedure can be applied to realize a pTDW with center position $z_0$, width $\Delta$, tilting attitude $\phi_{\mathrm{d}}$ and boundary condition $\theta_{\mathrm{d}}(\pi-\theta_{\mathrm{d}})$. First a short and strong enough field or current pulse is exerted to induce a wall around $z_0$ and after a transient process it finally becomes static in easy plane with Walker's profile. Then a series of ferromagnetic scanning tunneling microscope (STM) tips are placed along the wire axis with fixed tilting attitude $\Phi_0$ to produce a series of localized TMF pulsed. By arranging these tips with proper spacing and distance to strip, the envelope of these pulses is tuned to be the TMF profile in Eq. (\ref{Statics_Hperp_profile_v2}). The resulting static wall profile is the pTDW shown in Eq. (\ref{Statics_pTDW_profile}). When driving by axial field, since the transient process prior to traveling-wave mode is short (picoseconds), the STM tips can be arrange to move at the velocity in Eq. (\ref{Dynamics_velocity_v2}) so as to synchronize with the pTDW.

At last, our ``orientation-fixed" strategy can be generalized to the cases where pTDW motion is induced by spin-polarized currents, spin waves or temperature gradient, etc. Similar discussions can be performed to realized these pTDWs with clear boundaries. Magnetic nanostrips bearing with these walls would serve as proving ground for developing new-generation nanodevices with fascinating applications.

\section{Conclusions}

In this work, the ``orientation-fixed" TMF profiles are adopted to realize pTDW with arbitrary tilting attitude in biaxial magnetic nanostrips. After solving the LLG equation, unlike the classical Walker ansatz we obtain a pTDW with clear boundaries with adjacent domains and linear polar angle distribution inside wall region. More interestingly, the wall width can be freely tuned for specific usages. With TMF profile synchronized along with, these pTDWs can propagate along strip axis with considerably high velocity (well above that from the Walker ansatz) when driven by axial magnetic fields. These results should provide new insights in developing fascinating new-generation magnetic nanodevices based on DW propagations in nanostrips.

\vspace{6pt} 



\authorcontributions{Conceptualization, M.L. and J.L.; Methodology, M.L.; Validation, J.L.; Formal analysis, J.L.; Investigation, M.Y. and M.L.; Writing—original draft preparation, M.Y.; Writing—review and editing, M.L. and J.L.; Supervision, J.L.; Project Administration, J.L.; Funding Acquisition, J.L.}

\funding{This research was funded by the National Natural Science Foundation of China (Grants No. 11374088).}


\conflictsofinterest{The authors declare no conflict of interest.} 

\abbreviations{The following abbreviations are used in this manuscript:\\

\noindent 
\begin{tabular}{@{}ll}
DW   & Domain wall\\
TDW  & Transverse DW\\
pTDW & planar TDW\\
TMF  & Transverse magnetic field\\
UTMF & Uniform TMF\\
LLG  & Landau-Lifshitz-Gilbert\\
1D   & one-dimensional\\
1D-CCM & 1D collective coordinate model\\
1D-AEM & 1D asymptotic expansion method

\end{tabular}}




\reftitle{References}





\end{document}